\documentclass[
 reprint,showkeys,showpacs,preprintnumbers,
 amsmath,amssymb,
 aps,
 prl,
 longbibliography,
 lengthcheck,%
]{revtex4-1}

\usepackage{graphicx}
\usepackage{bm}
\usepackage{amsmath}
\usepackage[bookmarks=false]{hyperref}
\usepackage{dcolumn}

\begin{document}

\title{Characterizing vortex beam with angular-double-slit interference}

\author{Ruifeng Liu, Junling Long, Feiran Wang, Yunlong Wang, Pei Zhang,$^{\ast}$ Hong Gao, and Fuli Li}

\address{
MOE Key Laboratory for Nonequilibrium Synthesis and
Modulation of Condensed Matter, \\ Department of Applied Physics, Xi'an
Jiaotong University, Xi'an 710049, People's Republic of China
\\
$^*$Corresponding author: zhangpei@mail.ustc.edu.cn
}

\begin{abstract} The Fraunhofer diffraction intensity distribution of Laguerre-Gaussian beam is studied in an angular-double-slit interferometer. We demonstrate that the spiral phase structure of vortex light can be clearly revealed in this interference geometry, and it gives us an efficient way to distinguish different order of Laguerre-Gaussian beams. This angular-double-slit interference gives us a better understanding to the nature of orbital angular momentum and the interpretation of vortex beams interference phenomenon. 
\end{abstract}

\maketitle

It is well known that a photon of vortex light with an azimuthal phase distribution of the form $exp(il\varphi)$ carries a well defined orbital angular momentum (OAM) of $l\hbar$ \cite{Allen92,Yao11}, where $\varphi$ is the azimuthal angle and $l$ is an integer which describes the topological charge of vortex light. Because of the singularity on propagating axis, it results a dark point in the central of transverse intensity distribution. The OAM's degree of freedom is infinite, so it offers a good source for the quantum information process. It has been used in quantum computation \cite{Langford04,Nagali09,Zhang12}, quantum cryptography \cite{Terriza01} and high capacity of free space optical communication \cite{Gibson04,Wang12}. It is also can be used as optical tweezers \cite{Dholakia11} to control micro-object, spiral phase contrast imaging \cite{marte05} which can enhance the edge contrast and holographic ghost imaging \cite{Jack09}.

Characterizing the topological charge of an optical vortex beam attracts increasingly attention in recent years, and several methods have been proposed to sort and detect the OAM states. Generally, interference is a convenient way to analyze the phase structure of vortex light. Such as interfering the measured vortex beam with a uniform plane wave or it's mirror image \cite{Harris94}, in which the topological charge can be obtained by the number of forks or spiral petals. Some groups reveal the vortex phase profile by diffraction pattern with special mask. Berkhout \emph{et al.} \cite{Berkhout08} proved that a multiple pinhole diffracted system can be used to probe the OAM state of the measured optical vortex. Hickmann \emph{et al.} \cite{Hickmann10} pioneered a triangular aperture diffraction to measure the vortex phase structure. Beside these, single slit and double slit also have been proposed to analyze the topological change of vortex beam \cite{Sztul06,Guo09,Ferreira11,Mourka12}. Padgett's group has proposed two schemes to sort different OAM states under single photon level. One scheme was based on Mach-Zehnder interferometer with Dove prism \cite{Leach02,Leach04}, and the other scheme which converting OAM states into transverse momentum states \cite{Berkhout10}. However, most of these proposes are often involved with an unexpected complexity of interferometric patterns or a complicated interferometric experimental setup, which make them difficult to be executed in practice.

Due to the inherent spiral phase distribution $exp(il\varphi)$, we propose a scheme which can conveniently characterize the modulus and sign of the vortex beam's topological charge with angular-double-slit. This scheme is analogous to Young's double-slit pattern with constructive and destructive interference that depends of the phase difference between the double-slit. The angular-double-slit interference would give us a better understanding to the nature of OAM and the interpretation of vortex beams interference phenomenon.
\begin{figure}[htb]
\centerline{
\includegraphics[width=7.5cm]{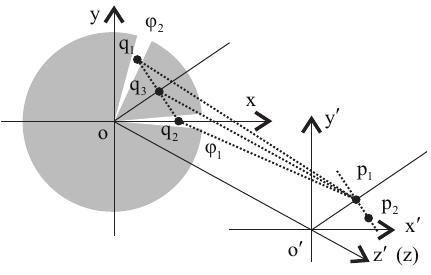}}
 \caption{The sketch of angular-double-slit interference.} \label{sketch}
\end{figure}

We start with a simple illumination of the proposed method theoretically. A schematic diagram of the scheme is shown in Fig.~\ref{sketch} with an angular-double-slit mask. $\varphi_{1}$ and $\varphi_{2}$ are two angular positions for two slits on the input plane. $q_{1}$, $q_{2}$ and $q_{3}$ are three points on the mask, and $oq_{3}$ is the angular bisector of $\angle q_{1}oq_{2}$. When a plane wave illuminates the angular-double-slit, a constructive interference pattern occurs at the $o'p_{1}$-axis where $o'p_{1}$ is parallel to $oq_{3}$ because of $|q_{1}p_{1}|=|q_{2}p_{1}|$.
A point $p_{2}$ deviating from $o'p_{1}$-axis will experience a constructive or destructive interference pattern according the optical path difference $|q_{1}p_{2}|-|q_{2}p_{2}|$.

If the illuminating light on the mask owns a spiral phase front $exp(il\varphi)$, the phase difference between $q_{1}p_{1}$ and $q_{2}p_{1}$ is:
\begin{align}
\Delta\phi=l\Delta\varphi+2\pi\frac{|q_{1}p_{1}|-|q_{2}p_{1}|}{\lambda},
\end{align}
where $\Delta\varphi=\varphi_{2}-\varphi_{1}$. If $\Delta\phi=N\pi$ (where $\lambda$ is the wavelength of the illuminating light) and $N$ is an even number, a constructive interference pattern occurs at the axis of $o'p_{1}$.
Once $N$ is an odd number, there will appear a destructive interference pattern at the axis of $o'p_{1}$. As shown in Fig.~\ref{sketch}, the angular-double-slit is symmetrical to $o'p_{1}$-axis,
so the interference pattern on $o'p_{1}$-axis is only determined by the phase difference $l(\varphi_{2}-\varphi_{1})$. For a given vortex beam with OAM number of $l$,
we will get a constructive or destructive interference pattern on $o'p_{1}$-axis depending on the angular difference of the angular-double-slit. For simplicity of operation, we can fix one slit at $\varphi_{1}=0$, and rotate the other slit $\varphi_{2}$ continuously. Thus,
the $o'p_{1}$-axis is at $\varphi_{2}/2$ direction, and we will obtain periodic constructive or destructive interference pattern at the $o'p_{1}$-axis. If the period is $2\varphi_{0}$, the topological charge of the vortex beam will be
determined by $l=\pi/\varphi_{0}$ except that a constructive interference pattern always occurs when $l=0$.

In the cylindrical coordinate system, Laguerre-Gaussian (LG) modes and Bessel beam are kinds of vortex beam which possessing OAM. LG modes are characterized by two mode indices. The radial structure is mainly determined by the radial mode index $p$ whereas the phase structure is described by the azimuthal mode index $l$. To simplify the numerical calculation, we set $p=0$ and the complex amplitude of LG modes is proportional to:
\begin{align}
E_{l}(r,\varphi)\propto exp(-\frac{r^{2}}{\omega^{2}})exp(-il\varphi),
\end{align}
where $\omega$ is the waist size of the beam. We first assume that the aperture angle and radius of angular-slit $\varphi_{1}$ in Fig.~\ref{sketch} is $\Delta\theta$ and $r_{0}$. As the single angular-slit can be approximately seen as a triangular aperture, the Fraunhofer diffracted field of this triangular aperture with LG modes can be expressed as:
\begin{align}
u(r',\varphi')&\propto \frac{r_{0}}{y'}e^{-i\pi(x'+y'\frac{\Delta\theta}{2})r_{0}}sinc(\pi(x'+y'\frac{\Delta\theta}{2})r_{0}) \nonumber\\
&-\frac{r_{0}}{y'}e^{-i\pi(x'-y'\frac{\Delta\theta}{2})r_{0}}sinc(\pi(x'-y'\frac{\Delta\theta}{2})r_{0}).
\end{align}
The diffracted pattern of the other angular-slit is $u(r',\varphi'+\Delta\varphi)$. In this way, we get the angular-double-slit interference complex amplitude:
\begin{align}
u_{l}(r',\varphi')= u(r',\varphi')+u(r',\varphi'+\Delta\varphi)e^{-il\Delta\varphi}
\end{align}

\begin{figure}[htb]
\centerline{
\includegraphics[width=7.5cm]{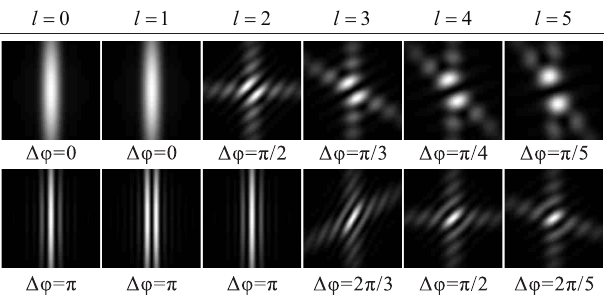}}
 \caption{Numerical simulation results of the angular-double-slit Fraunhofer diffracted patterns with different OAM states. $\Delta\varphi$ is angular difference of the angular-double-silt $\varphi_{2}-\varphi_{1}$.} \label{simulation}
\end{figure}
Fig.~\ref{simulation} shows the numerical simulation results of the angular-double-slit Fraunhofer diffracted patterns with different OAM states $l=0\sim5$. In the simulation process,
We rotate the second angular-single-slit and fix the first angle on the x-axis. The angular-single-slit width is $10^{\circ}$. When $l=0$, it always be a constructive interference pattern along the angular bisector of the angular-double-slit whatever the value of $\Delta \varphi$. It is analogous to angular-double-slit interference with plane wave. If the illuminating light owns an OAM $l=1$, there is a $\pi$ phase difference when $\Delta\varphi=\pi$ and a destructive interference can be observed at the angular bisector direction of the angular-double-slit. When $l=2$, the destructive and constructive interference pattern can be observed at the angular-double-slit bisector when $\Delta\varphi=\pi/2$ and $\Delta\varphi=\pi$, respectively. So it is obvious that the OAM states can be determined by the destructive and constructive interference pattern of angular-double-slit.


\begin{figure}[htb]
\centerline{
\includegraphics[width=7.5cm]{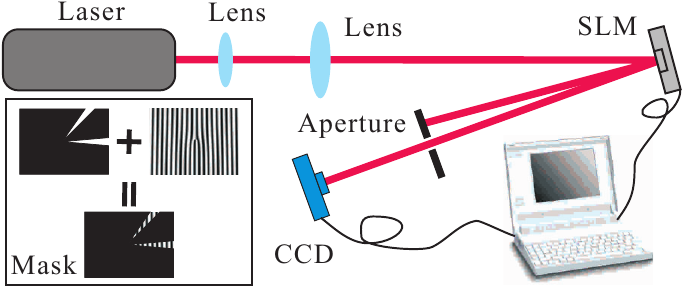}}
 \caption{The sketch of angular-double-slit Fraunhofer diffraction.} \label{setup}
\end{figure}

The experimental setup of angular-double-slit interference is shown in Fig.~\ref{setup}. The light emitted from the He-Ne laser is expanded with two lenses, and then illuminates a
computer hologram with controllable pixels written in an LCOS-SLM X10468 spatial light modulator (SLM). The inset of Fig.~\ref{setup} shows the mask written on SLM, and the mask is
a combination of an angular-double-slit and a hologram grating which generate higher-order LG modes. The first-order diffracted beam is chosen with an aperture, then a charge coupled
device (CCD) is used to read diffracted pattern. In the experimental process, one angular-slit is fixed on the x-direction, the other angular-slit is rotated around the center of crystal screen.

\begin{figure}[htb]
\centerline{
\includegraphics[width=7.5cm]{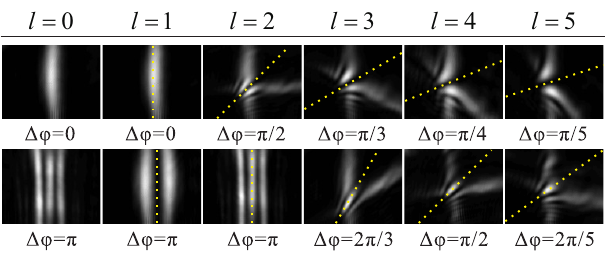}}
 \caption{Experimental results of the angular-double-slit Fraunhofer diffracted patterns with different OAM states. The dash line in the insets tilt $\Delta\varphi/2$ to horizontal direction.} \label{result1}
\end{figure}

Fig.~\ref{result1} shows some experimental results of the angular-double-slit interference. As analyzed above, a constructive interference pattern always can be obtained at the angular bisector direction of the angular-double-slit when the illuminating light is OAM $l=0$. The dash lines in Fig.~\ref{result1} denote the destructive and constructive interference direction for different OAM states and the inclination angle of the dash line in each inset is $\Delta\varphi/2$. From the results we know that, at the angular bisector direction of the angular-double-slit, a constructive or destructive interference pattern can be obtained in some special angular difference because the phase difference of the spiral phase profile $exp(il\varphi)$.

\begin{figure}[htb]
\centerline{
\includegraphics[width=7.5cm]{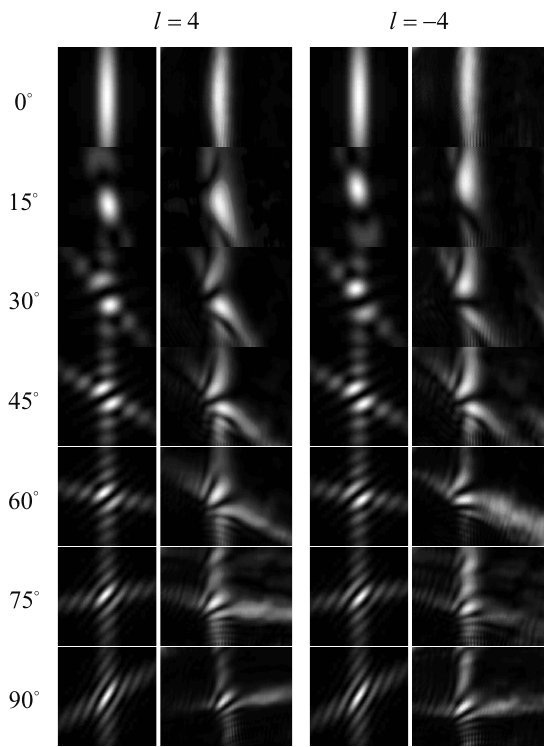}}
 \caption{Angular-double-slit Fraunhofer diffracted patterns with different OAM states $l=\pm 4$. The first and third column are numerical simulation, and the other are experimental results.} \label{result2}
\end{figure}

The detail of variation of the interference pattern along the intersection angle of the angular-double-slit is shown in Fig.~\ref{result2} with $l=\pm 4$. From the simulated and experimental results, we find the interference pattern have opposite variation tendency between the transition of the constructive and destructive interference for opposite sign of OAM states. It means that the constructive and destructive interference patterns move in opposite direction for OAM $+l$ and $-l$. Such as the process from $\Delta\varphi=0$ to $\Delta\varphi=90^{\circ}$, the constructive interference pattern moves from the bottom half to up half for $l=4$, while it moves from the up half to bottom half for $l=-4$. This result clearly reveals the spiral phase profile of vortex beam.

In conclusion, we demonstrate theoretically and experimentally that the interference pattern of an optical vortex after passing through an angular-double-slit can be used to characterize the the modulus and sign of the vortex beam's topological charge. With a unique intersection angle $\Delta\varphi=\varphi_{2}-\varphi_{1}$ of the angular-double-slit, a destructive or constructive interference pattern occurs at $\Delta\varphi/2$ direction, which can be used to determines the OAM states. The sign of OAM states can be obtained from the moving direction of the interference pattern. Our proposal does not need complicated interferometric setup and complicated interference patterns. Note that although previous works observed the interference fringes using multiple pinhole, triangular aperture, double- and single-slit experiments (in orthogonal coordinates, but not in the polar coordinates)\cite{Berkhout08,Hickmann10,Sztul06,Ferreira11}, our proposal offers us a better understanding to the spiral phase profile of vortex beam. For a more precise characterizing the topological charge of vortex beam, we can rotate the second slit a round to obtain the period of constructive or destructive interference pattern, and then determining the phase profile of vortex beam. We believe this proposal could be useful in the applications of quantum communication processing and astronomy with vortex beam.

This work is supported by the Fundamental Research Funds for the
Central Universities, Special Prophase Project on the National Basic
Research Program of China (Grant No. 2011CB311807), and the
National Natural Science Foundation of China (Grant Nos. 11004158,
11074198, 11174233 and 11074199).

\end{document}